\documentclass[%
 reprint,
 amsmath,amssymb,
 aps,
]{revtex4-2}

\usepackage{graphicx}
\usepackage{dcolumn}
\usepackage{bm}
\usepackage{bbm}
\usepackage{array}
\usepackage{multirow}

\usepackage{parskip}

\usepackage{tabularx}

\begin{document}

\preprint{APS/123-QED}

\title{Network lexical analysis of student integration of music and physics knowledge}

\author{W. Brian Lane$^{1,2}$}
 \email{Brian.Lane@unf.edu}
\author{Emily E. Pipkin$^{3}$}
\affiliation{%
 $^{1}$Department of Physics, University of North Florida\\
 $^{2}$Northeast Florida Center for STEM Education, University of North Florida\\
 $^{3}$Department of Biology, University of North Florida\\
 1 UNF Drive, Jacksonville, FL, 32224
}%

\date{\today}

\begin{abstract}

Introductory physics students benefit from learning to integrate new physics knowledge with prior knowledge related to their interests. Writing assignments that offer high epistemic agency can foster this integration, and provide rich artifacts for qualitative analysis. Physics education researchers are increasingly interested in studying such artifacts using computer-based textual analysis. In this paper, we use \textit{lexical methods} to look for evidence of knowledge integration in written assignments from a Physics of Music course. These assignments require students to communicate to an external audience a clear and cohesive narrative that synthesizes concepts from music and physics. To apply this methodology, we first establish a lexicon of music, physics, and integrated keywords that $N = 10$ students used across $6$ writing assignments. Then, we visualize this word usage in a series of network diagrams based on which keywords were used in proximity to each other in each assignment. We find that students most strongly connected ideas from music and physics using keywords related to the physical structure of sound (\textit{sound}, \textit{frequency}, and \textit{wave}). Applying these methods to each assignment prompt and its resulting student submissions, we find that, in assignments with a lower degree of epistemic agency (assessed qualitatively by our experience teaching the course), students primarily followed the example of integration modeled by the assignment prompt, but in assignments with a higher degree of epistemic agency, they explored music-physics integration to a greater degree. We use this analysis to offer a \textit{pedagogical contribution} to introductory physics courses centered around students' extant interests, a \textit{theoretical contribution} by illustrating a means of quantifying epistemic agency, and a \textit{methodological contribution} that complements textual analysis and offers the scalability sought in such approaches. 

\end{abstract}

\maketitle


\section{Introduction: Domain Knowledge Integration as a Subject of Lexical Analysis\label{sec:intro}}

An important challenge that physics educators face is supporting students in the process of integrating their new physics knowledge with existing knowledge from a subject that holds their interest. For many undergraduate students, these existing interests are embodied in their major, but for others they might be related to their career plans, hobbies, or communities. Regardless of the source, the opportunity for a student to explore the conceptual connections between physics and a domain of interest is an important act of \textit{epistemic agency}.

Epistemic agency is defined as ``the ways in which students, teachers, or social groups take on and distribute cognitive authority---that is, the responsibility for creating and evaluating new knowledge'' \cite{odden2023using}. When empowered as epistemic agents, students establish their own goals, determine their own methods for reaching those goals, and construct new knowledge in pursuit of those goals rather than reproducing established knowledge \cite{damcsa2010shared,hardy2020data}. Instructors can scaffold epistemic agency along different paths, opening up different aspects of the knowledge-building process depending on the activity or student population \cite{ko2019opening}. Epistemic agency is socially situated \cite{barton2010we}, such that a student might feel more agency within their domain of interest but less agency within a new context like physics. Students have been reported to prefer higher-agency activities, often attributing this preference to the freedom they experience \cite{kalender2021restructuring}.

Exploration of interdisciplinary concepts between physics and a domain of interest is an act of epistemic agency in that cognitive authority is invested in the student to implement their knowledge of the domain of interest while developing a new base of physics knowledge. This investment empowers the student to choose goals and methods based on their existing knowledge base and employ new physics concepts within that existing knowledge base. The knowledge constructed is new both as a development of physics concepts and as an extension of their existing interests. There is also the possibility that the student employs extant knowledge or develops interdisciplinary knowledge that is new to the instructor, heightening the authenticity of their work as an epistemic agent at the boundary of two domains.

While this integration is an exciting and personally fulfilling educational outcome, it can be difficult for instructors and researchers to find evidence of how students carry out this integration. For example, students might make conceptual connections that we do not know to ask about, and each student's background knowledge and interests are unique. Even if the instructor or researcher has a strong background in both physics and the domain of interest, differences between the instructor's and students' lived experiences in the domain of interest will invariably give a different shape to their knowledge and ways of thinking about the boundary between the two domains. However, we find that one can look for evidence of integration in how students connect ideas from the two domains in their writing assignments in a physics course.

While analyzing student writing has historically been prohibitively time-consuming, physics education research has seen a recent boom in successful applications of computational methods like natural language processing (NLP) to thematically categorize large sets of textual artifacts. These methods usually involve choosing between an unsupervised process or a supervised process \cite{odden2020thematic}. In an \textit{unsupervised process}, the computer produces a set of emergent topics represented across the corpus of textual artifacts to be checked and interpreted by the researchers. In a \textit{supervised process}, the researchers provide a priori topics with training examples for the computer to assign to each textual artifact in the corpus. An unsupervised process has the advantage of revealing unexpected patterns or groupings across the corpus, but these emergent topics might be difficult to interpret. A supervised process has the advantage of directly connecting with an existing framework or hypothesis, but it might leave real, unanticipated patterns unexplored.

The case presented here of interdisciplinary writing offers a previously unexplored methodological middle ground. Because each domain addressed in the corpus (in our case, music and physics) employs its own \textit{lexicon} of keywords (e.g., \textit{tempo}, \textit{pitch}, and \textit{rhythm} for music and \textit{velocity}, \textit{frequency}, and \textit{rate} for physics), researchers can create an a priori categorization scheme for disciplinary keywords without specifying what topics those keywords might address. Lexical categorization then provides a structure by which researchers can examine disciplinary integration in student writing across the topics discovered or verified using textual analysis.

In this study, we use lexical methods to examine a small sample of students' post-lab writing assignments (10 students, 6 writing assignments each) in a Physics of Music course. This analysis offers a \textit{pedagogical contribution} to introductory physics courses centered around students' extant interests, a \textit{theoretical contribution} to the study of epistemic agency, and a \textit{methodological contribution} to textual analysis.
Each of these contributions is based on one of the following research questions (RQs):

\textbf{RQ1} (pedagogical). How do music majors integrate their extant music terminology with recently acquired physics terminology in post-lab writing assignments?

\textbf{RQ2} (theoretical). How can the lexical method of categorizing disciplinary keywords quantify epistemic agency in the form of interdisciplinary integration?

\textbf{RQ3} (methodological). How can the lexical method of categorizing disciplinary keywords reveal underlying structure of written artifacts?

In Section \ref{sec:lit}, we review the literature on writing in the physics curriculum, particularly focusing on the writing formats of the Computational Essay and the Letter Home that were leveraged in this course's writing assignments to foster epistemic agency. In Section \ref{sec:textual_analysis}, we review examples of textual analysis from education research that highlight how lexical methods can support previous approaches. Lexical methods require an existing theoretical model to shape the lexicon, so in Section \ref{sec:model}, we summarize our theoretical model of epistemic agency and interdisciplinary integration that informs our lexical analysis. In Section \ref{sec:methods}, we describe the Physics of Music Course, identify our positioning and interests as authors, and outline how we used word counts and network analysis to analyze student use of music and physics terminology. In Section \ref{sec:results}, we use the results of this analysis to characterize student post-lab writing assignments in comparison with the assignment prompts. In Section \ref{sec:discussion}, we discuss how our results support this work's three contributions outlined above. Finally, in Section \ref{sec:limitations}, we discuss limitations of our study and in Section \ref{sec:conclusions} we outline future directions and present our conclusions.

\section{Writing Assignments Promote Epistemic Agency in Physics Education \label{sec:lit}}

Writing is a central practice in physics education, employed to advance various student learning goals. The AAPT Recommendations for the Undergraduate Physics Laboratory Curriculum encourages instructors to include learning outcomes related to ``present[ing] results and ideas with reasoned arguments supported by experimental evidence and utilizing appropriate and authentic written and verbal forms'' \cite{kozminski2014aapt}. Similarly, the AAPT Recommendations for Computational Physics in the Undergraduate Physics Curriculum encourages instructors to include learning outcomes related to ``prepar[ing] professional documentation and presentations'' \cite{behringer2017aapt}. This idea of writing as authentic professional expression supports the development of epistemic agency, which positions students as constructors of knowledge socialized into authentic practices.

Hoehn and Lewandowski \cite{hoehn2020framework} organize writing-related learning goals into three overlapping categories: writing as professionalization,  writing as communication, and writing to learn. Writing as professionalization involves helping students write like a scientist using ``forms authentic to the discipline'' \cite{kozminski2014aapt}. This category includes scientific argumentation because scientists must develop the skill of making a claim supported by evidence and reasoning \cite{mcneill2011claims,zwickl2013process,gillen2006criticism}. This category also includes the appropriate use of disciplinary jargon, as students learn to communicate physics ideas using physics terminology.

Writing as communication involves helping students write a clear and cohesive narrative, a skill that transcends an immediate scientific context \cite{hoehn2020incorporating}. Within this category, it is important for the final product to be understandable to a reader unfamiliar with the work \cite{magnifico2010writing}, a communication skill that science education may struggle with developing in students \cite{mccarthy2025we}. Because the purpose of a document often involves persuading the reader (to act in a certain way, or to accept the writer's conclusions), this category also includes scientific argumentation from the perspective of effectively communicating an argument.

Finally, writing to learn emphasizes how writing helps students master content, reflect on the learning process, and synthesize ideas that otherwise might be left disconnected in the student's mind \cite{reynolds2012writing,hand2004using,moskovitz2011inquiry,blakeslee1997activity}. In this perspective, a written artifact is not merely an output of a learning activity but a formative part of the learning process. For example, Buggé \cite{bugge2023improving} demonstrated that the revision process provides additional opportunities to develop scientific abilities and can improve student learning attitudes. Bylander and Gustafsson \cite{bylander2021improved} demonstrated increased learning gains after refining the requirements for a lab report assignment.

Each of these goals positions writing as an expression of epistemic agency: students adopting professional-like practices, students communicating findings to a designated audience, and students taking authority and responsibility for their own learning process.

For this study, two particularly relevant writing forms in physics education are the Computational Essay and the Letter Home. In a Computational Essay \cite{odden2025teaching,odden2019physics,odden2023using}, students weave together executable code, graphical results of computational modeling and data analysis, and explanatory text to ``make an argument, explain an analysis, or tell a story'' \cite{odden2025teaching}. Computational Essays are often written using the Jupyter notebook format \cite{kluyver2016jupyter}. Jupyter notebooks are made of code cells (executable segments of Python code that share a common memory) and markdown cells (lightweight html-based formatted text). Many physics instructors find the Jupyter notebook an effective environment for creating interactive computationally-integrated worksheets, modeling the process for students to create artifacts from computationally rich physics activities \cite{caballero2025integrating, lane2021analysis}.

The Computational Essay introduces ``authentic modes of communication within physics'' in a way that ``can be both communicative and reflective''  \cite{odden2025teaching}. In physics courses, Computational Essays are often used to foster epistemic agency \cite{odden2023using}. The Computational Essay, therefore, addresses all three of Hoehn and Lewandowski's goals for writing in physics: professionalization by practicing authentic writing modes, communication of a clear and cohesive narrative across the essay, and learning by helping students master and synthesize content \cite{hoehn2020framework}. 

The Letter Home \cite{lane2014letters,ramey2020comparative} was introduced as an alternative to the lab report. While much of the content of a Letter Home and a lab report might be the same (background information, experimental procedure, results, interpretations, and conclusions), the Letter Home requires the student to frame this content for a reader external to the course. The Letter Home was designed to address the lab report's shortcoming of requiring a student to write a narrative for their instructor: In authentic physics writing, one writes to an audience not familiar with the subject, while the lab report requires the student to communicate knowledge to the instructor who first delivered that knowledge to the student. 

Ramey et al. \cite{ramey2020comparative} found that Letters Home to non-expert readers demonstrated more authentic writing than Letters Home to expert readers. They otherwise observed similar outcomes between Letters Home and lab reports, meaning the informal nature of the Letter Home did not seem to sacrifice any of the outcomes one might typically look for in a lab report. The Letter Home addresses two of Hoehn and Lewandowski's goals for writing in physics: communication to an authentic audience and learning by deconstructing their understanding of content \cite{hoehn2020framework}. The Letter Home fosters epistemic agency in that it invests authority and responsibility for the student to choose a reader, assess their relevant knowledge, and then fashion the Letter Home appropriately.

These assignments' ability to foster epistemic agency, their focus on creating a coherent narrative, and their requirement to explicitly explain conceptual connections all align with our purpose of finding evidence of music-physics integration. In the Physics of Music course studied here, students were assigned to write \textit{Computational Letters}, a Computational Essay addressed and sent to an external reader like a Letter Home. We describe the Computational Letter as implemented in this course further in Section \ref{sec:course}. 

\section{Textual Analysis Would Benefit from Lexical Methods\label{sec:textual_analysis}}

Computer-based textual analysis techniques greatly expand the toolkit available to education researchers interested in studying large sets of textual data. For example, Berger and Toubia used NLP to analyze 40,000 college application essays to evaluate the volume of semantic ground covered (the breadth of domains from which ideas were drawn) and semantic speed (the size of conceptual jumps made between textually adjacent ideas) \cite{berger2024topography}. Their methodology relied on representing each essay as an ordered sequence of points in a high-dimensional latent semantic space, with each chunk of text represented as one point in the semantic space. By quantifying each essay's narrative in this way and comparing their semantic metrics against the students' subsequent performance in college, they were able to show that students who covered more semantic ground (high volume of diverse concepts) at a slower speed (smaller steps between adjacent ideas) tended to perform better throughout their academic careers, even when controlling for student-specific factors like high school GPA and test scores.

Within science education research, Odden and colleagues have used the unsupervised latent Dirichlet allocation (LDA) process to analyze themes across the \textit{Physics Education Research Conference Proceedings} \cite{odden2020thematic}, \textit{The Physics Teacher} and \textit{Physics Education} \cite{caramaschi2025analyzing}, and \textit{Science Education} \cite{odden2021has}. This application of LDA involved looking at the co-occurrence of words within a manuscript to identify latent topics across each corpus of articles. This process quantified the distribution of those topics across each journal and highlighted longitudinal shifts in topical focus across each journal's lifetime. As these themes are the result of an unsupervised process, they require interpretation by the researchers, who must review the keywords associated with each topic using their insider knowledge as physics education researchers. 

Similar processes are being applied to large datasets of student free responses, a type of data which has long presented a major time demand for physics education researchers. Campbell et al. carried out a benchmark test of IBM's Watson's capability of using supervised NLP to categorize explanations of student reasoning in short-answer problems using a priori codes \cite{campbell2024evaluating}. They found that categorizations to which Watson assigned high confidence scores tended to receive high agreement with human coders, but not items with low-confidence scores, ``suggesting that confidence score is correlated with increased labeling accuracy.'' They identified that Watson's categorizations particularly fell short for statements that could fit multiple categories or no categories. However, they found that small changes in answer phrasing to clarify student meaning reduced the number of incorrect categorizations. From this benchmark, they recommend that ``in datasets where human coders would like to use an `uncategorizable' code, it is important to develop a method to interpret Watson's scores as uncategorizable as well.'' They also conclude that human coding and checking will always be important in NLP approaches, both in preprocessing the data and interpreting the results.

Wilson et al. used NLP to classify open-ended responses to the Physics Measurement Questionnaire \cite{wilson2022classification}. They found that NLP categorized responses ``essentially as well as a pair of human coders'' but emphasized a frequent caution that NLP can propagate biases from humans and systems that shape the process. Wulff et al. used NLP to filter high-level reasoning elements from preservice teachers' written reflections \cite{wulff2023enhancing}. Using this process, they found they could identify differences in topics based on the teachers' disciplinary expertise. Bralin et al. compared outcomes from LDA and Non-negative Matrix Factorization (NMF) to analyze student essays from an introductory physics course \cite{bralin2023analysis}. They showed that error rates decreased with increasing number of topics, with LDA's error rate saturating after a sufficiently large number of topics and NMF's error rate decreasing toward 0 as the number of topics approached the number of essays in the dataset. Allen et al. used NLP to analyze scientific argumentation in students' writing, showing that scaffolded writing prompts tended to lead students to focus their writing around physics principles instead of procedural and surface-level features \cite{allen2025assessing}. Finally, Munsell et al. demonstrated how NLP could predict with 80\% accuracy whether students correctly solved a problem based on their written explanations, with a bias against correct scores (being labeled incorrect) \cite{munsell2021using}.

Tschisgale et al. describe a formal process of computational grounded theory \cite{tschisgale2023integrating,nelson2020computational} that consists of pattern detection (``leverag[ing] the power of computational techniques... for pattern detection in large datasets''), pattern refinement (``human researchers... add quality and depth to the quantity and breadth of [pattern detection]''), and pattern confirmation (``us[ing] computational techniques to test the extent to which the detected and refined patterns... hold throughout the whole dataset''). Reflecting on \cite{blei2012probabilistic} and \cite{rost2013representation}, they emphasize that ``the results of an analysis such as distributions of words over documents or syntactic networks do not speak for themselves'' and that ``when textual data is reduced to statistics in quantitative analyses, important and contextual factors may get lost or nonexisting relationships may be suggested.'' This challenge is comparable to computational modeling of physical phenomena, where it is often advised that the physicist cannot simply ``throw the problem into a computer'' and expect to receive an insightful solution. Instead, both the computational physicist and the textual analyst must employ a model based on existing knowledge of the problem, even in an unsupervised process. Tschisgale et al. demonstrate computational grounded theory using NLP techniques to classify students' descriptions of problem-solving approaches (pattern detection), then mapping the computer-identified topics onto known phases of problem solving (pattern refinement), which allows them to use both an a priori model and computational techniques to assess expert-like behavior in individual student texts (pattern confirmation). They also argue that these computational approaches to text analysis promote new avenues of replicating and extending previous results by researchers sharing their data and analysis scripts.

From this growing literature base, it seems that physics education researchers (and STEM education researchers more broadly) are keen to take up textual analysis techniques (primarily NLP) to finally realize the goal of representing and analyzing students' textual artifacts in a way that preserves student open expression (i.e., not constraining them to multiple-choice options) but maintains a manageable workload for the researcher. The primary challenge in implementing these approaches is that of \textit{interpretation} at the stage of deciding between supervised and unsupervised processes, preprocessing of textual data, and making sense of the process's results. We propose that \textit{lexical methods} can enhance the quality of this interpretation by providing a middle ground between supervised and unsupervised processes, providing a categorical layer for preprocessing the textual data at a lexical level, and making sense of the keywords that constitute the resulting thematic topics. Because these methods are applied at the scale of the dataset's \textit{lexicon} (instead of the individual textual artifacts), they maintain the scalability needed by researchers while increasing the insight available from the inerpretive acts.

Lexical methods ``rely on simple document-level word counts to accomplish tasks'' \cite{fesler2019text}. These methods ``can be applied when researchers know the types of terms that make up their construct of interest'' and can identify how those terms are appropriate for testing a particular theory \cite{fesler2019text}. Lexical methods require a corpus of documents, a lexicon of each word used across the corpus, and an appropriate unit of analysis for identifying where or how often each word in the lexicon appears. Because the lexicon simply identifies each unique word used across the corpus, it maintains a manageable scale regardless of the size of the corpus. For example, if one's corpus of interest was student lab reports about a pendulum experiment, students are likely to use a common set of words (e.g., \textit{pendulum}, \textit{energy}, and \textit{period}) regardless of whether the corpus includes 10 lab reports or 10,000 (e.g., it is unlikely the 9,999th lab report will include the word \textit{resistor}). A corpus can be preprocessed in short order by removing common words (e.g., \textit{the}, \textit{of}, \textit{are}) and categorizing the remaining salient words based on the model of interest to the researcher (e.g., scalar versus vector quantities, concepts versus lab equipment, or math terms versus physics terms).

In this paper, the types of terms that make our construct are music and physics keywords, and our theory is that students will attempt to integrate disciplinary knowledge between music and physics if afforded sufficient epistemic agency. Our lexicon is therefore composed of music and physics keywords, identified from the corpus of post-lab writing assignments the students submitted. We categorize these keywords as specific to the domain of music (e.g., \textit{timbre}), specific to the domain of physics (e.g., \textit{graph}), or an integration of the two (e.g., \textit{hertz}, which is used frequently in both domains). A simple count of when students use keywords from each category and how they co-occur within student writing can illustrate interdisciplinary thinking at the document level. To contrast this approach with NLP, an NLP method like LDA attempts to identify topics based on individual words that co-occur frequently (for example, \textit{potential}, \textit{kinetic}, and \textit{work} to identify a topic of energy). A lexical method begins with high-level word categories and quantifies how words from those categories are used together (for example, music and physics words being used together in a paragraph).

This approach can support the interpretive tasks required in textual analysis like NLP, whether integrated directly into the computational analysis or enacted in parallel to it. For both supervised and unsupervised processes, lexical categorizations can be used to refine or train the language model by modifying how words are embedded in the latent semantic space. For example, if one were measuring semantic distance like \cite{berger2024topography}, music keywords could be modeled as semantically close to each other in a cluster, as could physics keywords, with integrated keywords occupying a space equidistant from the music and physics clusters. This model refinement answers the call from \cite{tschisgale2023integrating,bamman2017natural} to increase NLP performance by training models on in-domain data, and addresses the call from \cite{campbell2024evaluating} to carefully attend to preprocessing of the textual data.

When interpreting the topics that result from unsupervised NLP, these lexical categorizations can help provide a mesoscopic layer of sensemaking between the keywords in a topic identified by the computer and the meaning that the researcher assigns to that topic. For example, in a literature analysis like \cite{odden2020thematic,caramaschi2025analyzing,odden2021has}, lexical categorizations could help distinguish keywords related to methodology, physics subjects, or institutional context. 

Tschisgale et al. identified a key limitation of their assumption ``that a sentence in a textual description always corresponds to exactly one theme. However, it is typically not uncommon that two or more themes are embedded within a single sentence'' \cite{tschisgale2023integrating}. Lexical methods allow researchers to identify multiple domains touched on by each unit of analysis, and quantifying the degree to which they do. Similarly, one can address their concern about statistical approaches losing contextual factors by encoding such contextual factors in the lexicon.

Finally, in answer to Tschisgale et al.'s call for researchers to share their data and analysis scripts \cite{tschisgale2023integrating}, a lexicon and its categorizations are straightforward to share. STEM education researchers working in conceptually similar domains could develop, share, and refine standardized lexicons with commonly used categorizations. Researchers could compare the effectiveness of using one shared lexicon versus another, much as we do with language models or conceptual inventories.

In this paper, we illustrate how lexical methods can reveal patterns of interdisciplinary integration in a small sample ($N = 10$ students writing $6$ assignments) of student writing with clearly distinguishable domain keywords. Because this sample is small and drawn from a course we are familiar with, we are able to make straightforward interpretations of the resulting structures in the data. However, the small sample size prevents us from also applying NLP directly in this study. Instead, we present this study as a proof-of-concept of the insights available from lexical methods that are of interest in NLP studies, with plans to integrate the two methodologies with a larger corpus in the future.

\section{Our Model: Epistemic Agency, Interdisciplinary Integration, and Word Use\label{sec:model}}

\begin{figure*}
    \centering
    \includegraphics[width=1\linewidth]{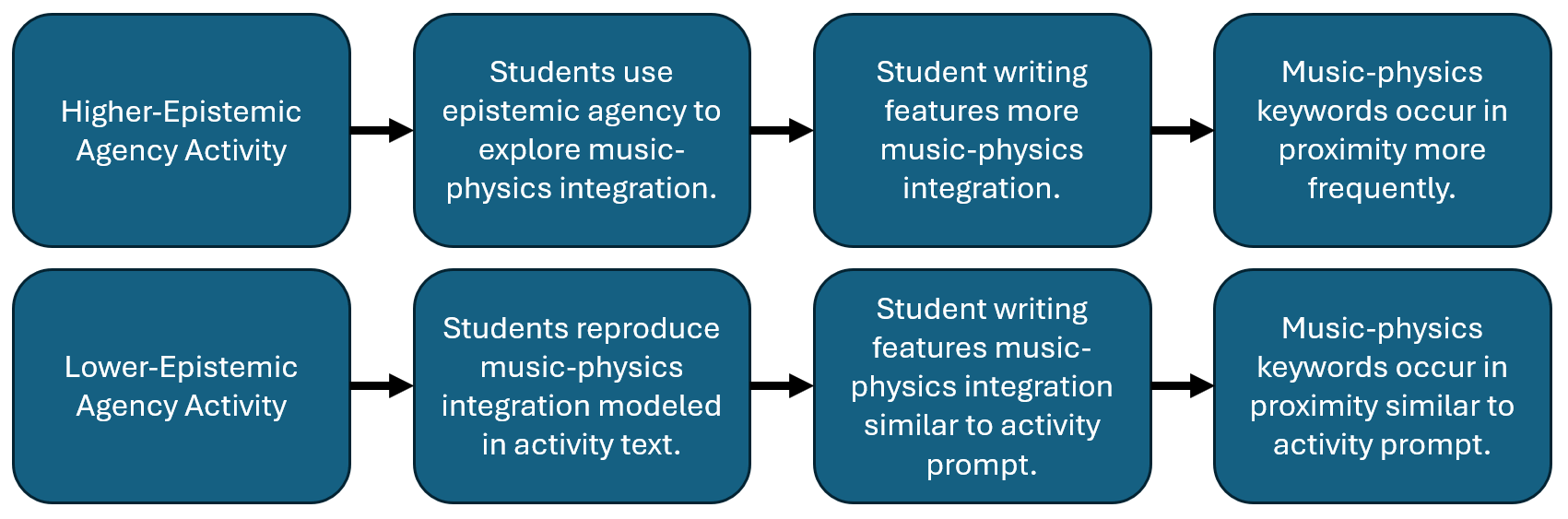}
    \caption{Visual depiction of our theoretical model. We expect that, in a Physics of Music course for music majors, lab activities that foster more epistemic agency will result in students' using that epistemic agency to explore music-physics integration. This integration should be observable in their writing as music and physics keywords co-occurring in proximity more frequently than in the activity prompt.}
    \label{fig:model}
\end{figure*}

As described above, lexical methods are applicable when researchers can associate categories of keywords with a particular theoretical model. For the proof-of-concept described in this paper, our categories comprise disciplinary keywords related to music, physics, and music-physics integration as employed in a Physics of Music course for music majors. We associate these categories with a model of epistemic agency in this Physics of Music course, depicted in Figure \ref{fig:model}. In this model, a post-lab writing assignment with high epistemic agency will exhibit more disciplinary integration, since that is an important part of these music majors' taking authority and responsibility for their learning process (writing to learn). This epistemic agency is fostered by assignment features that encourage the students to adopt authentic physics practices (writing as professionalization) to communicate the application of physics to music to a designated audience (writing as communication). We look for evidence of disciplinary integration in the students' word use. Following the assumptions of textual analysis methods, we assume that disciplinary keywords co-occurring in proximity to each other indicates disciplinary integration, such that the more music and physics keywords that co-occur in proximity, the more disciplinary integration is featured in a unit of text. 

However, these written artifacts are shaped by the lab activity they are written in response to. As we will demonstrate with a lexical analysis of the lab activity prompts, some lab activities in Physics of Music can be more predisposed to disciplinary integration than others. For example, the classic physics lab of forming standing waves on a string is incredibly relevant to music but does not require significant music knowledge to carry out or to write a report about. On the other hand, a lab activity in which students characterize the spectral profiles of different instruments directly integrates knowledge from the music and physics domains. Therefore, we evaluate an assignment's epistemic agency by comparing students' integration of disciplinary terminology with the integration already present in the lab activity prompt. 

To summarize, we are proposing a means of organizing semantic concepts by their a priori domain of origin, thereby quantifying the interdisciplinarity of student writing as a metric of epistemic agency. Word use is our data, integration is our metric, and assignment epistemic agency is our theoretical construct of interest. To our knowledge, this is the first time lexical methods have been employed to quantify epistemic agency in this way, and our results will verify the appropriateness of this measure.

\section{Context and Methodology \label{sec:methods}}

Here we discuss relevant information about the Physics of Music course under study, our backgrounds and roles as authors in this study, and the major steps of our lexical methodology.

\subsection{The Physics of Music Course\label{sec:course}}
This Physics of Music course studied here is a general education physics course designed for all students and specifically tailored to the background and interests of music majors. Although the course was open to all students who met the college algebra prerequisite, in the fall 2023 section that we study, only music majors ($N=10$) enrolled in the course, with a breadth of concentrations (instrumentalists and vocalists studying music education, music performance, and music technology). As further described in \cite{lane2025computing}, this integrated lecture-laboratory course was designed around the following student learning outcomes (SLOs):
\begin{enumerate}
    \item Demonstrate an understanding of the nature of sound, including overtone series/harmonic series.
    \item Determine the ratio of frequencies for various intervals. 
    \item Be able to decompose multiple sounds into their component frequencies for various instrument tone qualities and study its connections to timbre.
    \item Describe the acoustics of various instruments. 
    \item Identify the speed at which sounds travel in various media and its consequences on the acoustics of different performance venues. 
    \item Measure sound characteristics in the context of health and safety guidelines.
    \item Describe the acoustics of spaces.
\end{enumerate}

\begin{table*}
    \centering
    \begin{tabular}
    {|p{0.16\textwidth}|p{0.16\textwidth}|p{0.16\textwidth}|p{0.16\textwidth}|p{0.16\textwidth}|p{0.16\textwidth}|} \hline
         &  Lab Activity 1 Standing Transverse Waves
&  Lab Activity 2 Standing Sound Waves
&  Lab Activity 3 Frequency Spectrum of Similar Instruments
&  Lab Activity 4 Frequency Spectrum of Different Instruments
& Lab Activity 5 Performance Hall Acoustics
\\ \hline
         SLO 1 Nature of sound
&  \multicolumn{1}{c|}{X}
&  \multicolumn{1}{c|}{X}
&  \multicolumn{1}{c|}{X}
&  \multicolumn{1}{c|}{X}
& 

\\
         SLO 2 Ratio of frequencies
&  \multicolumn{1}{c|}{X}
&  
&  \multicolumn{1}{c|}{X}
&  \multicolumn{1}{c|}{X}
& 

\\
         SLO 3 Decompose sound into component frequencies
&  
&  
&  \multicolumn{1}{c|}{X}
&  \multicolumn{1}{c|}{X}
& 

\\
         SLO 4 Acoustics of instruments
&  
&  
&  \multicolumn{1}{c|}{X}
&  \multicolumn{1}{c|}{X}
& 

\\
         SLO 5 Speed of sound
&  \multicolumn{1}{c|}{X}
&  
&  
&  
& 

\multicolumn{1}{c|}{X}
\\
         SLO 6 Measure sound characteristics 
&  \multicolumn{1}{c|}{X}
&  
&  
&  
& 

\multicolumn{1}{c|}{X}
\\
         SLO 7 Acoustics of spaces
&  
&  
&  
&  
& 

\multicolumn{1}{c|}{X}
\\ \hline
    \end{tabular}
    \caption{Student learning outcomes (SLOs) and lab activities in Physics of Music. Lab Activities 1 and 2 are fairly straightforward explorations of the physical nature of sound like one might carry out in a traditional introductory physics course. Lab Activities 3 and 4 utilized students' instruments to create spectrographs to study similarities and differences in timbre. Lab Activity 5 characterized the acoustics of the recital hall where the course took place. Lab Activity 6 (not listed here) required student groups to choose a topic to investigate (therefore addressing differing SLOs for each group).}
    \label{tab:labs}
\end{table*}

Students worked toward these outcomes in a series of six laboratory activities (outlined in Table \ref{tab:labs}) that each spanned two weeks. The first five activities were structured as guided inquiry labs. The final activity was an open-ended project in which student groups chose a topic to study, specifically designed to promote students' epistemic agency \cite{odden2023using,holmes2020traditional,kalender2021restructuring}. Each activity was presented to students in a starter Jupyter notebook (SJ) with instructional text and starter Python code for the students to use and develop \cite{oleynik2019scientific}. These SJs were hosted on Google Colab to enable students to easily make their own copy, edit them, and share with their selected reader.

At the conclusion of each lab activity, students were required to write a Computational Letter (CL), a Computational Essay written to an external reader. To create a CL, students were required to replace the SJ's instructional text (color-coded red to easily identify) with explanatory descriptions of what they did to complete the lab activity and what they learned in the process. Writing to an external reader required each student to consider their reader's unique background in music and physics and frame their writing accordingly. In their CLs' markdown cells, they were expected to explain the purpose and function of each code cell and use the outputs of these code cells to explain and support their conclusions. 

\begin{table}
    \centering
    \begin{tabular}{|p{0.35\textwidth}|c|c|} \hline
        The Computational Letter is written in a Jupyter notebook. All red text has been replaced by the students' narrative. & Yes & No  \\ \hline
        The Computational Letter makes one or more claims connected to each Student Learning outcome (SLO) at the beginning of the activity. & Yes & No \\ \hline
         The Computational Letter presents evidence from the lab activity that support its claims. & Yes & No \\ \hline
         The Computational Letter describes how the evidence was collected in the lab activity. & Yes & No\\ \hline
         The Computational Letter makes a reasoned argument that connects the evidence to their claims. & Yes & No \\ \hline
    \end{tabular}
    \caption{Single-point rubric used to grade each CL draft. ``Red text'' refers to the instructional text students were tasked with replacing with explanations for their reader.}
    \label{tab:rubric}
\end{table}

Upon completing their first draft of each CL, each student submitted the draft to the instructor through the course's on-line learning management system (Canvas). The instructor would then grade the submitted draft with the single-point specifications rubric depicted in Table \ref{tab:rubric}. This rubric was designed to emphasize student reflection on the learning process and scientific reasoning skills applicable to each activity. The draft's numerical score was not yet entered into the course grade, allowing the student to revise and resubmit their drafts until they were satisfied with their numerical score. The instructor supported this revision by allowing time in class for student-initiated conversations about rubric feedback and potential actions. Once satisfied with their revisions, the student emailed the link to the CL (usually with explanatory directions to open and read the letter) to their chosen external reader with the instructor in the message's CC field. The instructor then entered the CL's most recent numerical score into the course grade. 

Based on the course's goals and student audience, the CL format and revision process emphasize writing to learn (promoting content mastery, reflection, and synthesis), writing as communication (presenting a clear and cohesive narrative with a persuasive argument to an external reader), and writing as professionalization (describing authentic music and physics practices). This assignment format and revision process is particularly helpful in promoting music-physics disciplinary integration. The Computational Essay format requires weaving disciplinary ideas together in a coherent narrative around modeling and data analysis. Writing to an external reader requires explicitly connecting disciplinary concepts for an audience who might hold significant disciplinary knowledge in music or physics, but not likely both. The revision process gave the instructor opportunities to prompt students to further explain their thinking about physics as related to music.

It is in the text of these CLs that we look for evidence of how students integrated their extant music knowledge with their newly acquired physics knowledge by examining their use of terminology from each discipline. 

\subsection{Author Backgrounds and Information \label{sec:authors}}
Both authors have a background in music performance. The second author (EEP) is an undergraduate biology major and the first author (WBL) is a physics education researcher who taught the Physics of Music course under study here. We undertook this study out of an interest to explore an important aspect of the learning process in introductory physics. We initially hypothesized that the results might demonstrate how the students learned to integrate music and physics concepts over the timeline of the semester, and found our results to be surprising but sensible after completing the analysis described in Section \ref{sec:results}.

EEP led the analytical process described below, using her experience performing music and taking introductory physics to delineate between music and physics keywords. WBL used his experience teaching the Physics of Music course and his backgrounds in music and physics to provide clarification during the analytical process and to interrogate keyword categorizations. We found these complementary roles to support an objective but instructionally informed analysis of the qualitative data collected.

\subsection{Extracting and Categorizing Keywords from the Computational Letters}

\begin{figure*}
    \centering
    \includegraphics[width=1\linewidth]{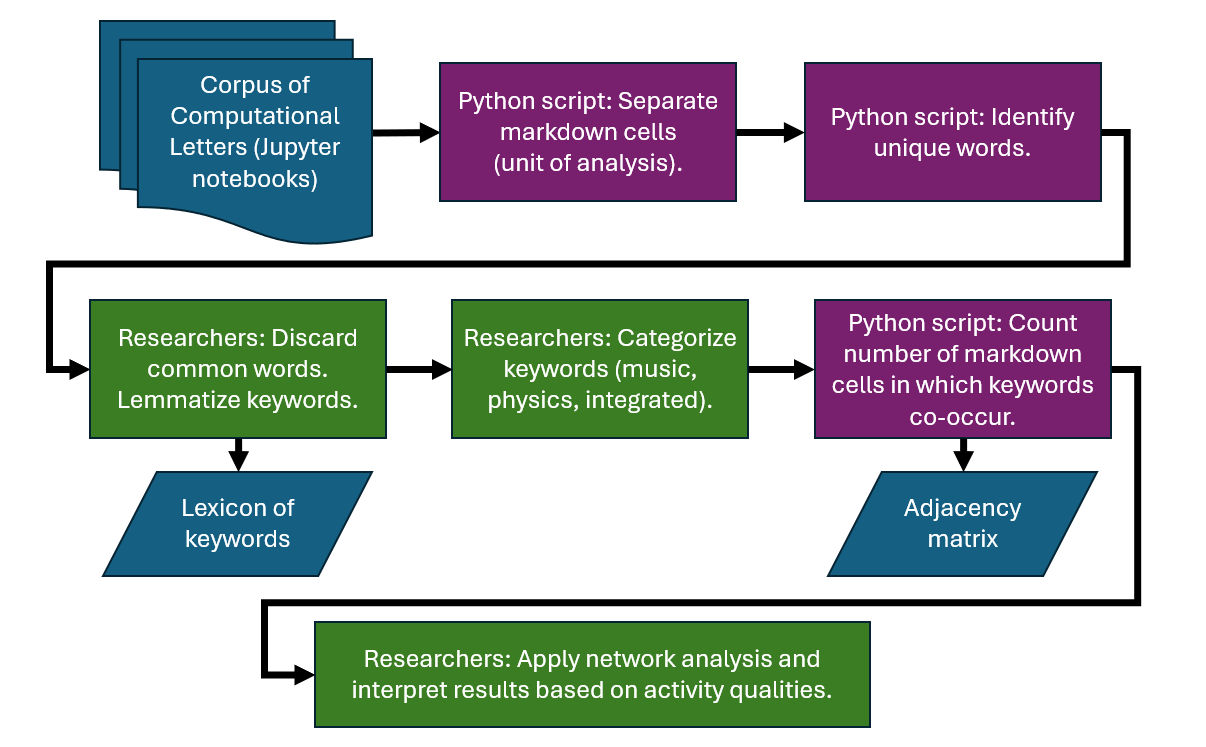}
    \caption{Visual depiction of our lexical method, with data in blue, computer-automated tasks in purple, and researcher tasks in green. We began with a corpus of Computational Letters in the format of Jupyter notebooks. We ran a Python script to separate these notebooks into their individual markdown cells to use as our unit of analysis. This script further identified the set of unique words within this corpus, which we simplified by removing common words and lemmatizing the remaining keywords (e.g., combining ``frequency'' with ``frequencies''). We then categorized the keywords as associated with the domain of music, the domain of physics, or an integration of the two. Finally, we used our Python script to count the number of co-occurrences of keywords to create an adjacency matrix and carry out the network analysis described in Section \ref{sec:network_analysis}. The Python scripts used in this approach are available at \cite{git}.}
    \label{fig:method}
\end{figure*}

In studying how students demonstrated integration of music and physics knowledge, we sought to identify how they used disciplinary terminology in proximity to each other within each CL. Our methodology here follows the example of Fesler et al. \cite{fesler2019text} described above and is illustrated in Figure \ref{fig:method}. In our case, the types of terms that make our construct is music and physics \textit{keywords}.

We use ``keyword'' here to mean a term that has a specific, commonly understood meaning in the discipline of music or physics. This definition is similar to Cash, Finkelstein, and Hoehn's criteria for ``scientific vocabulary'' as used by children: ``words used in a science context that are not common in daily conversation... or in a way that differs from how they are used colloquially'' \cite{Cash2024writing}. 
For example, \textit{timbre} is a keyword from music and \textit{interference} is a keyword from physics. Integrating knowledge of these two disciplinary concepts might look like, ``An instrument's timbre is created by the interference of the overtones it produces.'' Music and physics also use some keywords in common, such as \textit{sound} or \textit{frequency}, with overlapping meaning.

Lexical methods require a lexicon of terms, a corpus of documents, and an appropriate unit of analysis. In our case, the lexicon is composed of music and physics keywords and the corpus is the CLs the students submitted. Choosing the unit of analysis requires some consideration, since we are interested in when students used two keywords in proximity to each other. Clearly, an entire CL is too large of a unit of analysis: A student likely knows they need to refer to certain keywords in an assignment but might not relate them extensively. For example, they might use music keywords only in the CL's introduction and physics keywords only in the analysis. However, an individual sentence might be too small: A student might connect two keywords in a chain of reasoning that spans multiple sentences. Fortunately, the Jupyter notebook provides a compromise that is easily parsable with a computer, the markdown cell. When a student writes in a Jupyter notebook, they must do so within a given markdown cell at a logical place within the notebook. Although not a requirement, students in this course typically placed a single markdown cell between consecutive code cells, explaining how the steps carried out in each code cell build on each other and advance the study they describe. Therefore, for our study, the markdown cell captures a cohesive, small nugget of student reasoning, and we assume that when a student used two keywords within a markdown cell, they were intentionally integrating them in some way.

To create a lexicon for this study, we developed a Python script to extract each individual markdown cell from each CL to track our unit of analysis. The script then identified approximately 3500 unique words used across all CLs in the course. Next, we assembled our lexicon by identifying disciplinary keywords. In this process, we discarded common words like conjunctions or pronouns. We then combined conjugations and misspellings of each keyword (a process called lemmatization).

We categorized each keyword as related to music, physics, or an integration of the two; the full list of categories is available in the lexicon at \cite{git}. While some keywords clearly had roots in a particular discipline (such as \textit{timbre} from music or \textit{graph} from physics), we did not categorize keywords strictly based on disciplinary definitions, but rather the way the students collectively used them across the CLs. We attended to what they were describing in the lab activity, or how the keyword functioned in their overall argument. This attention to context also helped us distinguish between the disciplinary use of a keyword (``A \textit{wave} has a period and amplitude'') and everyday uses (``I \textit{waved} hello to my friend'').  

The second author (EEP) made the first pass over all keywords to assign initial categories or flag unclear keywords for discussion. The first author (WBL) made a subsequent pass to review the categories, discuss unclear keywords, and question potential errors. While most of the keyword categories were straightforward to identify and consensus was reached right away, we found we needed to discuss around one out of every five keywords. Some of these discussions were resolved quickly by noting that some keywords could be combined (for example, \textit{guitar}, \textit{trumpet}, etc. were all combined under \textit{instrument}) while others required reviewing how students used a keyword across the corpus.

For example, one of the first keywords we needed to review was \textit{hertz} (the unit of frequency), which we observed was used with both physics and music connotations. When discussing how to categorize \textit{hertz}, we considered how one student, in CL4, compared two spectrographs as follows: ``The violin had a lot more peaks than the oboe. But each division of the peaks were in multiples of 440 Hz (the harmonic intervals of the note).'' Here, the student associates 440 Hz (as seen on their spectrograph) with a specific note, as did another student comparing an instrument and a voice: ``The note both being sung and played was A4 or 440 Hz.'' Equating a particular hertz value with a note seems to indicate that it is used both as a music term and a physics term. Therefore, we categorized \textit{hertz} as an integrated keyword.

Another keyword that required further review was \textit{acoustics}, which we found to be used in both music and physics contexts. The students seemed to use this keyword either as a music-exclusive term when discussing an audience member's experience of a performance, or as a bridge between physics and music. For example, one student wrote in their CL5, ``Every area intended for music or entertainment has been specifically designed with acoustics in mind... to provide the audience with the finest possible listening experience.'' This use of \textit{acoustics} specifically refers to the design of a performance venue for an audience. Later in the same CL, this student wrote, ``We had to take decibel (amplitude) level readings at roughly half of the recital hall's seats in order to gauge the space's acoustics.'' Here, the student operationalizes \textit{acoustics} in terms of amplitude measured in units of decibels as a function of position, while still in the context of attendees sitting in a performance venue. Therefore, we categorized \textit{acoustics} as an integrated keyword.

While many of our keyword discussions involved nuanced review and resulted in a categorization of integrated, we also found that we needed to discuss important features of the course context, since the first author (WBL) taught the course while the second author (EEP) led the keyword analysis. For example, EEP inquired about the students' use of the keyword \textit{pulse}, which could, in principle, refer to something like tempo. However, WBL reviewed the students' use of \textit{pulse} and clarified that this word was used exclusively to refer to a pulse sent down a string in Lab Activity 1, such that we categorized \textit{pulse} as a physics keyword.

During our review discussions, we found that each keyword could fit clearly into one category (music, physics, or integrated) based on the students' overall usage across the CLs. We did not need to categorize words differently based on individual assignments (e.g., ``physics'' in one assignment but ``music'' in another). For example, while \textit{measure} carries a different meaning within the context of music and physics, our review confirmed that students never used \textit{measure} in the music sense in these CLs.


\subsection{Network Analysis\label{sec:network_analysis}}

Once we identified music and physics keywords, we extended our Python script to identify which keywords were used in each markdown cell of the SJs and CLs. As established in our theoretical model in Section \ref{sec:model}, we included the SJs in this process to help us understand how each lab activity modeled music-physics integration to compare with the students' writing in the CLs. The Python script counts the number of markdown cells with which each keyword was used within each SJ and CL and creates an \textit{adjacency matrix} of the keywords. In this matrix, each off-diagonal element $a^{\alpha}_{ij}$ ($i\neq j$) counts the number of markdown cells in a notebook (or set of notebooks) $\alpha$ that contain both keyword $i$ and keyword $j$, and each diagonal element $a^{\alpha}_{ii}$ counts the number of markdown cells that contain keyword $i$. 

For example, one student's CL3 opened with the following introduction in the first markdown cell. We denote each music keyword from our lexicon in bold, each physics keyword in italics, and each integrated keyword in bold italics: ``The purpose of this \textit{lab} was to \textit{analyze} the \textbf{harmonics} of two similar \textbf{instruments}, and dissect their \textbf{\textit{spectrum}} \textit{graphs} so we can attempt to copy the \textbf{timbre}. We chose to record an `A' on the \textbf{cello} and the double \textbf{bass}, two very similar \textbf{instruments}, and then studied the \textbf{overtone} structure of each.'' This brief markdown cell contains 3 music keywords (\textit{cello} and \textit{bass} are combined into \textit{instrument}), 3 physics keywords, and 1 integrated keyword, therefore contributing to ${7 \choose 2} = 21$ off-diagonal elements in the adjacency matrix. These off-diagonal elements numerically encode the student making interdisciplinary connections between the data visualization of a spectrograph to an instrument's timbre composed of overtones. With each subsequent markdown cell that the student further discusses this integration, these off-diagonal elements increase.

To obtain a high-level view of the contents of each SJ and CL, we define a keyword category frequency $f^{C\alpha}$ for each category $C = M,P,I$ (music, physics, integrated) as
\begin{equation}
    f^{C\alpha} = \frac{\sum_{i \in C} a^{\alpha}_{ii}}{\sum_i a^{\alpha}_{ii}}.
\end{equation}

A category with a higher frequency has more keywords mentioned per markdown cell in the SJ or CL $\alpha$. 

However, the mere presence of keywords from a category within the SJ does not guarantee that the notebook models the integration of keywords from music and physics. For example, one could construct a SJ that uses many different music keywords in a motivational introduction but then pivots into heavy use of physics terminology when describing the lab activity. Therefore, we also explore a high-level view of how each SJ or CL integrates keywords from each category by defining a category integration frequency $f^{CC^\prime\alpha}$
\begin{equation}
    f^{CC^\prime\alpha} = \frac{\sum_{i \in C, j \in C^\prime, i \neq j} a^{\alpha}_{ij}}{\sum_{i \neq j} a^{\alpha}_{ij}}.
\end{equation}

In our earlier example of the student's opening markdown cell in CL3, there are $3 \times 3 = 9$ connections made between a music keyword and a physics keyword, another ${3 \choose 2} = 3$ connections made between pairs of music keywords, $3$ connections made between pairs of physics keywords, $3$ connections made between an integrated keyword and a music keyword, and $3$ connections made between an integrated keyword and a physics keyword (totaling to the $21$ off-diagonal elements counted earlier). We created bar graphs to show keyword category frequency (number of markdown cells in which keywords from the music, physics, or integrated categories were used) and the category integration frequency (number of markdown cells in which each pair of words from any two categories was used).

We then used the Python library networkx to create networks from these adjacency matrices, in which each keyword is represented as a node. The node's diameter is proportional to the number of markdown cells that contain the keyword, and the node's color represents the keyword's category (music, physics, or integrated). Two nodes are connected by an edge if the two keywords occurred in the same markdown cell (i.e., if their $a^{\alpha}_{ij}>0$), and the edge's thickness is proportional to the number of markdown cells in which they occur together $a^{\alpha}_{ij}$.

These network diagrams allow us to visually surmise the integration of music and physics keywords. A network diagram with few strong connections between music and physics keywords indicates a lack of integration, while a network diagram with many strong connections between music and physics keywords indicates strong integration. We can quantify such connections by evaluating each node's betweenness centrality. Such an evaluation helps us identify which concepts an instructor could emphasize to support disciplinary integration. To evaluate a node's connecting role across the network, we examine its subset betweenness centrality.

A node's betweenness centrality $b_{i}^{\alpha}$ \cite{barthelemy2004betweenness} is a fractional count of the number of geodesics (shortest paths) between any two nodes in the network that node $i$ lies along:

\begin{equation}
    b_{i}^{\alpha} = \frac{\sum_{j \neq k \neq i} g_{jk}^{\alpha i}}{\sum_{j \neq k} g_{jk}^{\alpha}}, \label{eqn:betweenness}
\end{equation}

where $g_{jk}^{\alpha}$ is the number of geodesics between nodes $j$ and $k$ in network $\alpha$ and $g_{jk}^{\alpha i}$ is the number of geodesics between nodes $j$ and $k$ that pass through node $i$. In our network, the distance between two connected nodes is evaluated as the inverse edge weight $1/a^{\alpha}_{ij}$. Subset betweenness modifies this calculation by considering how often a keyword mediates between one subset of nodes and another. In our case, we consider subset betweenness defined between the set of music keywords $M$ and the set of physics keywords $P$:

\begin{equation}
    b_{i}^{\alpha, M\leftrightarrow P} = \frac{\sum_{j \in M, k \in P} g_{jk}^{\alpha i}}{\sum_{j \in M, k \in P} g_{jk}^{\alpha}}, \label{eqn:betweenness}
\end{equation}

This subset betweenness therefore allows us to evaluate which keywords students most often used to integrate their music and physics knowledge in the CLs.

\section{Results \label{sec:results}}

In this section, we present the results of our lexical analysis from the Physics of Music SJs and CLs. First, we use keyword category frequency and category integration frequency to characterize the contents of each SJ. Starting with the SJs helps us examine how each activity modeled music-physics integration for the students. Then, we characterize the students' CLs using the same frequency metrics, allowing us to compare the students' writing against the SJ's model. Finally, we more thoroughly examine the CLs' contents through network diagrams and music-physics subset betweenness.

\subsection{Characterizing Starter Jupyter Notebooks}
Before we examine the content of the students' submitted CLs, we find it helpful to characterize the SJ provided at the start of each lab activity. These SJs included guiding instructions, minimally working code \cite{lunk2025student}, and questions for students to answer. Students were expected to replace all starter text with their own writing to their external reader and to expand the code cells to carry out necessary analytical processes. Characterizing the SJs helps us understand how each lab activity modeled music-physics integration for the students.

\begin{figure*}
    \centering
    \includegraphics[width=1\linewidth]{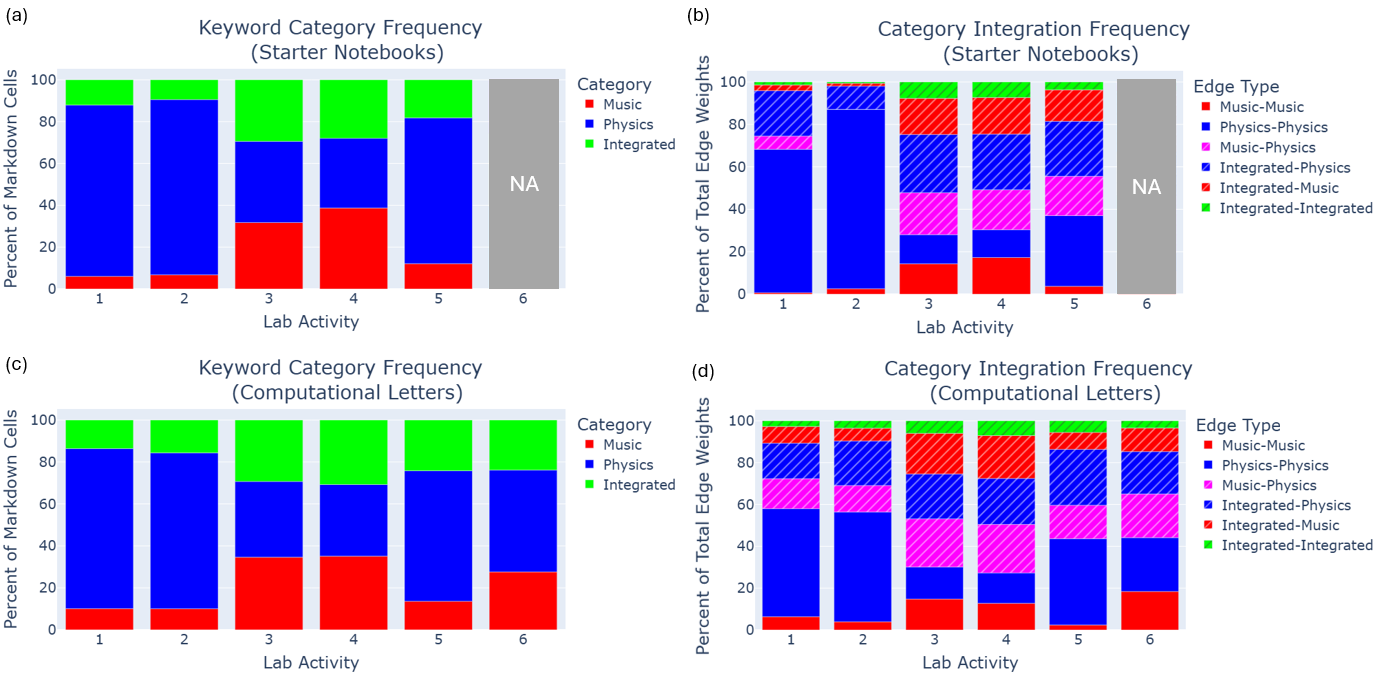}
    \caption{Content analysis of the starter Jupyter notebooks (a and b) and computational letters (c and d). The graphs of keyword category frequency (a and c) show the distribution of categories as a percentage of the total number of keywords counted using the markdown cell as the unit of analysis. The category integration frequency (b and d) shows the distribution of edges between keywords of different categories as a percentage of the total edge weight. Edges between nodes solely within music or physics are represented by solid bars and edges that connect music and physics concepts or involve at least one integrated concept are represented by hatched bars.}
    \label{fig:CL-content}
\end{figure*}

Figure \ref{fig:CL-content}(a) shows each SJ's keyword category frequency, the number of markdown cells in which keywords from the music, physics, or integrated categories were used. We omit the SJ for the final lab activity since it required students to choose a topic and develop their own process for studying that topic. The SJ for the final lab activity contained no procedural instructions, and instead simply presented an outline of the activity's timeline for completion and suggested topics. The distribution of category frequencies shows that the SJs for Lab Activities 1, 2, and 5 were dominated by physics keywords, while the SJs for Lab Activities 3 and 4 have greater representation of music and integrated keywords.

In Figure \ref{fig:CL-content}(b) we present each SJ's category integration frequency, the number of markdown cells in which each pair of words from two categories are used. We mark these frequencies with a solid bar if the two categories do not represent integration (two words from music or two words from physics in the same markdown cell) and add a stripe pattern if the two categories represent integration (one word from music and one word from physics in the same markdown cell, or at least one word from integrated). This analysis reveals that SJ1 and SJ2 feature a high frequency of physics keywords being used in the same markdown cells, with few instances of music and physics keywords being used together. We interpret this to mean that these SJs indeed focus on physics content, as suggested by their keyword category frequencies in Figure \ref{fig:CL-content}(a). On the other hand, SJ5 includes more music-physics integration than SJ1 and SJ2, even though Figure \ref{fig:CL-content}(a) shows that they have somewhat comparable keyword category frequencies. The category integration frequencies for SJ3 and SJ4 show that these SJs exhibit more integration between music and physics than the others.

These observations make sense in light of the activities' topics: Lab Activities 1 and 2 teach essentially the same standing waves content that one might encounter in a standard introductory physics course. While they have additional musical context added than one might typically present in a standard course, they do not specifically leverage students' music knowledge. Lab Activities 3 and 4 heavily use the students' own instruments in the lab procedures and required them to set up audio recordings in spaces appropriate for those instruments. Finally, the acoustical concepts involved in Lab Activity 5 represent a strong overlap in music and physics knowledge, but the activity itself does not specifically leverage music knowledge.

\subsection{Counts of Keywords and Categories}
\begin{figure*}
    \centering
    \includegraphics[width=1\linewidth]{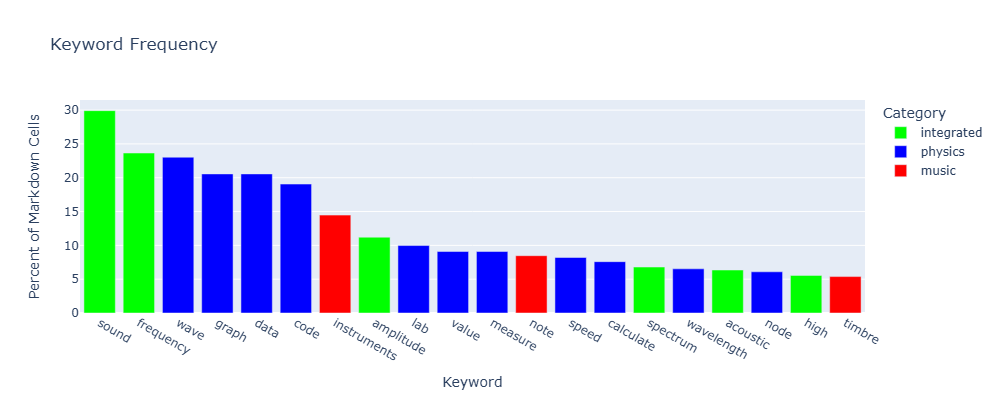}
    \caption{Distribution of keywords across all CLs by percentage of markdown cells they were used in.}
    \label{fig:keyword-count}
\end{figure*}

In Figure \ref{fig:keyword-count}, we show the frequency (number of markdown cells) of the 20 most frequently used keywords across all CLs. Choosing a cutoff of the top 20 allows us to examine the breakdown between our three categories with at least three keywords per category. The color of each bar indicates the category of the corresponding keyword (music, physics, or integrated). The most commonly found keyword, \textit{sound}, was used in 30\% of markdown cells and is an integrated keyword. The following 5 keywords (\textit{frequency}, \textit{wave}, \textit{graph}, \textit{data}, and \textit{code}) occur with similar frequency (around 20\%). Most of these are physics keywords. The high frequency of these keywords makes sense: \textit{Wave} is a fundamental concept when studying physics applications to music. \textit{Graph} and \textit{data} are particularly salient for these post-lab writing assignments, and the high frequency of \textit{code} reflects the computationally integrated nature of the lab activities.

The first music keyword, \textit{instruments}, occurs in fewer than 15\% of markdown cells (half as often as \textit{sound}). Students specifically used their instruments as part of Lab Activities 3 and 4, and discussed applications to instrument properties in CLs 1, 2, and 6. The remaining keyword frequencies drop off steadily as the frequency rank decreases. These 20 most frequently used keywords are primarily in the integrated or physics categories, with little representation of music-specific keywords. 

Examining the distribution of keywords by CL, we find variation in the relative frequency of music, physics, and integrated keywords. Figure \ref{fig:CL-content}(c) shows the keyword category frequency for each CL, intended to be compared with the same quantities for the SJs in \ref{fig:CL-content}(a). Overall, the keyword category frequency (c) for CLs 1-5 follows the pattern of the associated SJs (a). It seems that students generally followed the lead of the SJs, not introducing much additional terminology from any category. The content for CL6 shows slightly more physics keywords than music keywords, but the overall distribution is more similar to CLs 3 and 4, which had more music terminology present than the others.

We also find it important to consider how the students integrated music and physics keywords, which we explore in the next subsection.

\subsection{Music-Physics Keyword Integration\label{sec:integration}}

\begin{figure*}
    \centering
    \includegraphics[width=1\linewidth]{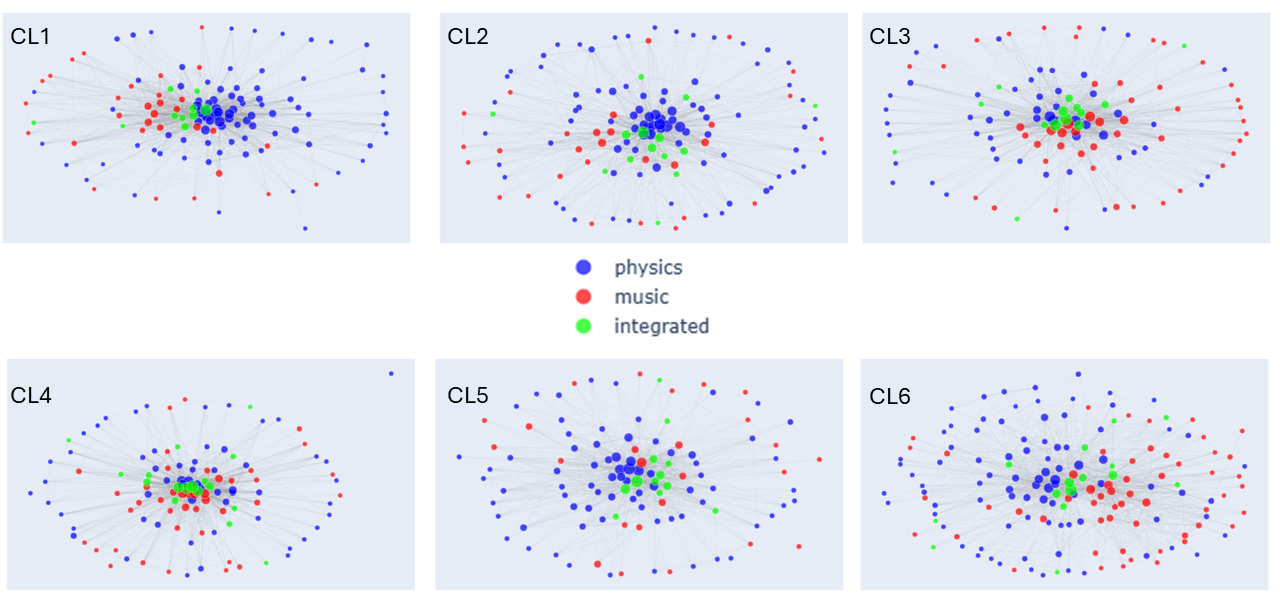}
    \caption{Network diagrams representing the keyword connections in each computational letter. Each node represents a keyword, color-coded by disciplinary category and sized by the number of markdown cells they were used in. Edges represent two keywords being used in the same markdown cell, with thickness proportional to the number of markdown cells.}
    \label{fig:networks}
\end{figure*}

The network diagrams in Figure \ref{fig:networks} give insight into how students integrated their disciplinary terminology across the six CLs. In these diagrams, each keyword is represented by a node (circle) whose diameter is proportional to the number of markdown cells the keyword was used in and whose color indicates the keyword's category. Two nodes are connected by an edge (line) if their keywords were used in the same markdown cell, with the edge's thickness proportional to the number of markdown cells in which the pair of keywords were used. The code we developed to create these network diagrams is available at \cite{git}, allowing the reader to view an interactive version of each network diagram where each node's keyword can be identified. However, to answer our research questions, it is sufficient to examine how the keyword categories feature in each network without specifying individual keywords. 

The network diagrams use the Kamada-Kawai layout \cite{kamada1989algorithm} in which the length of an edge is determined by a spring-force model that minimizes the total potential energy of the system. This layout means more well-connected nodes tend to occur closer together near the center of our diagrams while more isolated nodes tend to occur in the periphery. This representation allows us to notice how the network diagrams for CL1, CL2, and CL5 primarily centralize nodes from the physics category with only a few nodes from the integrated category in the center. On the other hand, the network diagrams for CL3 and CL4 tend to have nodes from all three categories near the center. CL6's structure is more diffused, reflecting how each student group chose a different topic to investigate for this lab activity, and therefore the class used a broader variety of terminology than they did in the first five lab activities.

To systematically study these networks, we examine the distribution of edges by the categories they connect and the music-physics subset betweeenness for the 20 keywords presented in Figure \ref{fig:keyword-count}.

\subsubsection{Category Integration Frequencies\label{sec:cl1_example}}

Figure \ref{fig:CL-content}(d) shows the category integration frequency for each CL, intended to be compared with the same quantities for the SJs in \ref{fig:CL-content}(b). We see a mostly similar pattern as in Figure \ref{fig:CL-content}(b): CLs 1, 3, 4, and 5 generally have a similar distribution to that of their SJs, with CLs 1 and 5 having more connections within the physics category and CLs 3 and 4 having both more integrated connections and more connections within the music category. And as with the keyword category frequencies, the distribution for CL6 looks more like the distributions for CLs 3 and 4.

For example, one student opened their CL1 with the following markdown cell (\textbf{music}, \textit{physics}, and \textbf{\textit{integrated}} keywords are formatted as above): ``In my \textit{Physics} of \textbf{Music} class, I got to learn about the study of \textit{wavelengths} and how they correspond with \textbf{\textit{frequencies}}. Since you don’t have much of a background on these things, let me catch you up. Every \textbf{note} we produce either out of our mouths or a noise from an object, we create a \textbf{\textit{sound}} with a \textbf{\textit{frequency}}. The \textbf{pitches} are \textit{wavelengths}, the \textbf{\textit{higher}} the \textbf{\textit{frequency}} (\textbf{\textit{higher}} \textbf{pitch}), the faster the \textit{speed} of the wave, and the \textbf{\textit{lower}} the \textbf{\textit{frequency}} (\textbf{\textit{lower}} \textbf{pitch}), the slower the \textit{wave} \textit{speed}. In this activity, my partner and I studied the \textit{wave} \textit{motion} of a string by using a \textit{meter} stick, stop watch, and slow \textit{motion} video.'' While this introduction connects many music and physics keywords, just two markdown cells later, when the student is describing their work in the activity, music terminology is almost entirely absent: ``We needed to determine the \textbf{\textit{frequency}} of our standing \textit{wave}, which is the number of \textit{wave} \textit{oscillations} per second (the number of times my partner moved his arm to create a consistent \textit{pulse}). We needed to determine our \textit{wavelength}, which for the first wave was the total length of the string \textit{divided} by the number of \textit{antinodes} (\textit{hills} and \textit{valleys}). This came to be roughly 162.7 \textit{centimeters}. We computed it for two other \textit{waves} as well, and noticed that as the \textit{antinodes} \textit{increase}, the \textit{wavelength} \textit{decreases}.'' This markdown cell reads like a lab report focused on explaining a physics-heavy procedure with little outlook toward relevance to music.

However, CL2's distribution of category integration in Figure \ref{fig:CL-content}(d) shows notable differences compared to SJ2's distribution in Figure \ref{fig:CL-content}(b). The physics-physics portion (representing markdown cells with each pair of physics keywords) is smaller by about one-third, with music-physics integration much larger (15\% compared to nearly 0\% in the SJ), and all counts of keyword pairs with at least one integrated keyword (music-integrated, physics-integrated, and integrated-integrated) are larger. Even though the overall integration distribution for CL2 looks similar to CL1 and CL5, it features more music-physics integration than the original SJs does. We interpret this difference based on unique features of Lab Activity 2 in Section \ref{sec:discussion}.

This unique difference between SJ2 and CL2 prompts the question of what content the students added to their CLs that produced this level of music-physics integration. We find that the highest-weight music-physics edges in the CL2 network that were not found in the SJ2 network are \textit{area}-\textit{venue}, \textit{lab}-\textit{amphitheater}, and \textit{wave}-\textit{performance}. These edges all seem to suggest a discussion of applications to performance venues.

\subsubsection{Music-Physics Subset Betweenness}

\begin{figure*}
    \centering
    \includegraphics[width=1\linewidth]{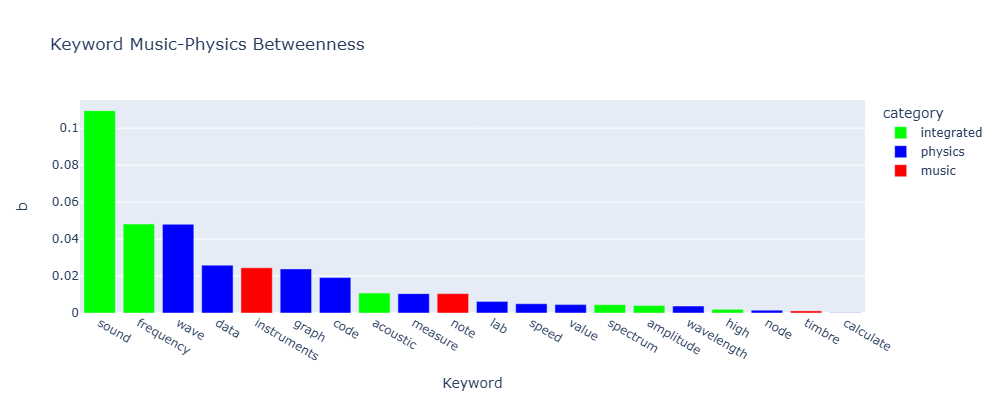}
    \caption{Subset betweenness values for the 20 most frequently used keywords. Subset betweenness quantifies how often each keyword connects two keywords from the music and physics categories.}
    \label{fig:music-physics-bewteenness}
\end{figure*}

Returning our attention to the full corpus of CLs, we can use the keywords' subset betweenness centrality to evaluate which keywords most strongly integrate the music and physics categories. Figure \ref{fig:music-physics-bewteenness} shows the music-physics betweenness for the 20 highest-frequency keywords. These values represent the fraction of shortest paths between each music node and each physics node that pass through the given node. Path length is computed using the inverse edge weight. This quantity allows us to identify how often these keywords help connect concepts unique to the music and physics domains.

By far, the keyword with the largest music-physics subset betweenness value is the integrated keyword \textit{sound}. It is found along approximately 11\% of the shortest paths between music and physics keywords, a rate of more than double the next-highest music-physics betweenness value, which is nearly a tie between \textit{frequency} and \textit{wave}. We contrast this large difference with the comparatively smaller difference in frequency of usage that \textit{sound} has with \textit{frequency} and \textit{wave} in Figure \ref{fig:keyword-count}. This simpler frequency count places these three keywords on closer-to-equal footing than music-physics betweenness, which highlights the more central role that \textit{sound} plays in the network of keywords. It seems that, while \textit{sound} might be used nearly as often as related concepts, it much more strongly connects the music and physics disciplines. 

Taking these three keywords together in the context of Physics of Music, \textit{sound}, \textit{frequency}, and \textit{wave} (two integrated keywords and one physics keyword) all refer to the physical structure of sound. It seems that this overarching concept, certainly foundational to the subject matter, consistently served an integrating role in the students' writing.

After \textit{wave} in Figure \ref{fig:music-physics-bewteenness}, the betweenness values drop sharply, with only \textit{data}, \textit{instruments}, and \textit{graph} (two physics keywords and one music keyword) rising above 2\%.

\section{Discussion \label{sec:discussion}}

These findings accomplish our stated purpose of offering a \textit{pedagogical contribution} to introductory physics courses centered around students' extant interests, a \textit{theoretical contribution} to the study of epistemic agency, and a \textit{methodological contribution} to textual analysis of students' written artifacts. Here, we discuss these contributions and answer their associated research questions.

\subsection{Pedagogical Insights into Student Interdisciplinary Exploration}

Here, we answer RQ1: How do music majors integrate their extant music terminology with recently acquired physics terminology in post-lab writing assignments? We observed that students integrated prior music knowledge (from their major) with new physics knowledge (from a Physics of Music course) based on a few key concepts (sound, waves, and frequency).

Reviewing Figure \ref{fig:keyword-count}, we note that the most-used keyword, \textit{sound}, is from the integrated category, being a physics phenomenon that is fundamental to the practice of music. However, of the next 19 highest-frequency keywords, 11 (more than half) are physics-specific, 4 are integrated, and only 3 are music-specific. This speaks to an overall dominance of physics-related terminology in the students' writing, which we also see in Figure \ref{fig:networks}. For a course whose goal is to promote integration of music and physics concepts, one might hope to see more equity between the terminology used in the two disciplines. At the same time, we also have to acknowledge that Physics of Music is structured as a physics course that requires students to learn new physics concepts and practices, and the students' writing might appropriately focus on the new knowledge gained.

Figure \ref{fig:music-physics-bewteenness} shows that \textit{sound}, in addition to being found in the most markdown cells, is the strongest connection point in the network between music keywords and physics keywords, having by far the greatest music-physics subset betweenness. \textit{Frequency} and \textit{wave} have the next-greatest subset betweenness values, indicating that focusing on the physical structure of sound might hold the greatest potential for prompting music majors to integrate music and physics concepts.

Interestingly, \textit{wavelength} has a much lower subset betweenness than \textit{frequency}, even though these are complementary properties that physicists often group together. For example, a physicist might say, ``If you know the wavelength, it's straightforward to find the frequency,'' or, ``Frequency and wavelength uniquely determine each other for a given speed of sound.'' On the other hand, these music majors used the keyword \textit{frequency} much more often than they used \textit{wavelength}, and they more strongly used it to connect music and physics ideas. We conclude this difference indicates that \textit{frequency} holds a much more central place in their thinking. Indeed, we categorized \textit{frequency} as an integrated keyword but categorized \textit{wavelength} as a physics keyword because these students (like most musicians) often referred to notes by their frequency (for example, ``A4 is 440 Hz'') but almost never by their wavelength.

We note that for many of these students, the physics keywords included in their CLs represent new knowledge that they have become familiar with and are shaping their thinking around. Although we did not assess students' incoming science knowledge, conversations indicated that most of them had not invested significant coursework in scientific subjects beyond the essential requirements in high shool. Two students in the class identified as music technology majors, which afforded them a bit more facility with the computational aspects of the course, but even these students found that they were exploring these practices at a deeper level than they had before. Overall, though, the CLs with higher degrees of integration (particularly the open-ended CL6) represent the students' perceptions of how this new knowledge applied to them personally.

These results support the use of epistemic agency as a pedagogical tool to promote student knowledge integration between physics and a domain of interest. Students need epistemic agency to pursue integration through inquiry and reflection, and it seems likely that students would want to use epistemic agency to integrate topics of interest to them. Even in a course in which the students' interests have little overlap with each other, fostering epistemic agency can support their varied interests.

\subsection{Quantifying Epistemic Agency through Interdisciplinary Integration}

When we first undertook this study, we initially hypothesized that, as students become more comfortable with physics concepts and practices, one might hope to see more music-physics integration as the semester progressed. We reasoned that, since disciplinary integration is a higher-order application of learned concepts, we would see more music-physics integration as students became more familiar with the physics concepts. Looking at the progression from CL1 through CL4 in Figures \ref{fig:CL-content}(c), \ref{fig:CL-content}(d), and \ref{fig:networks}, we see what could be evidence of this development. The networks of keywords for CL1 and CL2 (written in the first month of the semester) are more physics-centered, and they have more physics-physics edges than any other type. Then, networks of keywords for CL3 and CL4 (written in the second month of the semester) show music and physics sharing center stage with much more integrated edge types
. If the Physics of Music course had ended here, we might conclude that the students' integration of music and physics had simply increased as a function of time spent in the course, likely corresponding to greater familiarity with the physics content.

However, moving to CL5, the trend seems to be reversed: Music keywords have been decentralized in the network. There are fewer integrated edges than in CL3 and CL4
. If the transition from CL1 and CL2 to CL3 and CL4 represented students learning how to integrate their music and physics knowledge, does the transition to CL5 represent an unlearning of integration? And what, then, should we make of the increased integration in CL6?

This unexpected result is what led us to consider the construct of epistemic agency and formulate RQ2: How can the lexical method of categorizing disciplinary keywords quantify epistemic agency in the form of interdisciplinary integration? To answer this question, we compare the lexical analysis for each SJ with its corresponding CLs. For Lab Activities 1, 3, 4, and 5, we observe an overall match between each SJ 
and the students' use of keywords in their CLs: If the SJ heavily integrated music and physics keywords across its markdown cells, the students' CLs would heavily integrate music and physics as well. However, if the SJ centered physics keywords and marginalized music keywords, the students' CLs would focus on physics, as well. 
From this comparison, we conclude that, in Lab Activities 1, 3, 4, and 5, students were largely following the example of music-physics integration laid out in the SJs.

However, CL2 violates this pattern: SJ2's network has an even greater concentration (around $80\%$) of physics-physics edges, making it a heavily physics-focused SJ. In contrast, CL2 has many more music-physics edges than SJ2. Given the matching pattern shown by the other SJs and CLs, we might expect the physics-physics edge percentage to be higher for CL2 and the music-physics edge percentage to be lower.

We explain this difference by reflecting on the unique nature of Lab Activity 2: This activity involved setting up a single-frequency sound source in an outdoor walkway of uniform width with concrete walls to either side, thereby creating a two-dimensional standing wave pattern around the chamber. Students used sidewalk chalk to mark the locations of nodes and antinodes, producing a square grid pattern. This process perhaps most directly involved students in exploring the nature of sound, in an embodied process in which they were directly measuring the phenomenon with their ears, feet, and hands. The instructor and students recorded the locations of these markings in a single cloud-shared spreadsheet, then returned to the classroom to analyze the data together. Analysis involved graphing the marked points and creating a grid of lines to mark the patterns, then using the grid spacing to determine the sound's wavelength and then calculate the speed of sound.

Since the instructor had never conducted this activity before and knew that the process would need to be developed on the fly after data was collected, he wrote very few prompts in the SJ for the data analysis process. As a result, the instructor and students collaboratively developed the analysis process, identifying problems and making adjustments together. In other words, a significant portion of the activity was not pre-written in SJ2.

These differences amount to Lab Activity 2 having a greater degree of epistemic agency than Lab Activities 1, 3, 4, and 5. As described in Section \ref{sec:lit}, this process distributed cognitive authority across the instructor and students, as they collected data together and co-developed the analytical process. This distribution invested responsibility in the students to create knowledge (i.e., ``Where do I hear the sound the loudest and softest? How will we visualize this data? What can we learn from it?'') and evaluate this knowledge (i.e., ``This marking looks out of place. How do we know if this process worked? What standard value can we compare our value for the speed of sound to?''). We conclude that if the students were more involved in creating and guiding the processes for data collection and analysis in Lab Activity 2, this higher epistemic agency afforded by CL2 would result in a higher level of music-physics integration than we might expect from SJ2.

Our analysis of the edges introduced to CL2 supports this explanation. The highest-weight music-physics edges found in CL2 but absent in SJ2 include considerations of a performance venue, an important application of the standing wave concept developed in this lab activity. It therefore seems that, when presented with a ``gap'' in the activity instructions, students filled this gap with a discussion of music-physics integration.

This interpretation is supported by our earlier example of a student's introduction to their CL1 in Section \ref{sec:cl1_example}. The student's introductory markdown cell interwove music, physics, and integrated keywords under the premise of ``catching up'' their reader (a high-epistemic agency task). However, their markdown cell that explained the process of determining frequency and wavelength (a low-epistemic agency task) used physics keywords almost exclusively, following the procedure modeled in SJ1.

We also see evidence of higher epistemic agency producing greater disciplinary integration from the data for CL6, for which each group chose their own topic to investigate and explored which methods seemed appropriate for the task. As a result, SJ6 does not contain instructions aside from the general process and deadline for the end of the semester, which is why we do not include CL6 in Figure \ref{fig:CL-content}(a) and (b). This ``blank slate'' process has a high degree of epistemic agency, and leads to a higher degree of music-physics integration. Therefore, we conclude that students tended to use greater epistemic agency to explore music-physics integration, in agreement with our theoretical model (Section \ref{sec:model}).

This analysis validates our claim that an assignment's epistemic agency can be quantified using student integration of keywords from physics and a subject of interest to students. We find that assignments with higher epistemic agency in this course were directly associated with more integration when compared with the assignment prompt. A prevalence of disciplinary integration in the student artifacts implies evidence of epistemic agency in the assignment.

\subsection{Alternative Explanations for Integration Patterns}

Before answering our last research question, we consider alternative explanations to our results that do not involve epistemic agency. We developed the explanation above for the differences in integration across the CLs based on the first author's experiences teaching the Physics of Music course and grading the CLs, the second author's experience reviewing the CLs for context in how the keywords were used, and the construct of epistemic agency, which the course and these assignments were designed to promote. We now critique this explanation by considering alternative explanations, and argue why they do not fully explain the data.

A simple explanation might be that there was not much music content in Lab Activities 1, 2, and 5 to begin with, and therefore little for the students to write about. Similarly, it might be the case that the students simply mimicked the verbiage used in SJs 1 through 5. This explanation would imply that our construct of knowledge integration amounts to reading too much into the data. This is certainly a valid concern. However, this explanation is not supported by the contrast between SJ2 and CL2, where the students filled in the gaps in the instructional documentation with music-physics integration. Similarly, it seems that their ``default'' mode of writing for CL6 (which had no starter content for them to follow) includes a good deal of integration. It seems that simply presenting a physics activity that we as instructors know is music-relevant does not generate as much integration as one might hope without explicit room for epistemic agency.

Another simple explanation might be that we are observing effects of the time of semester when these activities took place. No doubt these students experienced varying pressures throughout the 15-week term (exams in other courses, performances, course registration opening), which might dispose them to being more engaged in the process in Lab Activities 3 and 4 and less in the others. Such external pressures are always important to keep in mind both as researchers and instructors. However, variation of external circumstances does not explain how the integration patterns in the CLs mostly align with those in the SJs. Additionally, due dates in this course were flexible, and students expressed appreciation for how this accommodated for external pressures. Therefore, it does not seem likely that these competing pressures were a significant factor in student engagement.

\subsection{Lexical Methods as a Complement to Textual Analysis}

Finally, we answer RQ3: How can the lexical method of categorizing disciplinary keywords reveal underlying structure of written artifacts? As discussed in Section \ref{sec:textual_analysis}, scalable analysis of student written artifacts is of growing interest to STEM education researchers. A primary goal of this methodology is identifying topical themes, trends, and structures in student writing that help us make sense of student thinking at an aggregate level while preserving students' freedom in written expression. This study illustrates how even broad categorization of a lexicon at the scale of each word's domain of origin can highlight differences in student writing based on assignment structure. We were able to characterize music-physics integration in assignment prompts and student writing using network analysis. This approach ties directly to how language models embed words in a semantic vector space, and how textual analysis uses word proximity as a metric to determine topics.

This sort of structural insight is of broader interest to textual analysis studies, which typically either require the researcher to already know what topics to tell the computer to identify (a supervised process), or require the researcher to hand over topic determination to the computer and interpret those topics afterwards (an unsupervised process). Lexical methods can complement and contribute to these processes in STEM education research by highlighting high-level conceptual structures that can help the researcher determine an approach, and by supporting the researcher's interpretive tasks.

For a supervised process, one could use lexical categories to develop codes at a broader scale before refining those codes to focus on more specific topics. For example, a supervised NLP technique could identify text as ``music-focused,'' ``physics-focused,'' or ``music-physics integrated'' before looking for specific topics within music or physics. Our examples of students' markdown cells demonstrate how the researcher can use keyword categorization to easily check and interpret this broad-scale coding.

For an unsupervised process, one could use lexical categories to refine the language model being used based on the study. For a corpus like ours, one could ensure that music keywords were semantically near each other and physics keywords were semantically near each other. Another option is to manually add to the semantic space one additional dimension for each domain. For example, if $\vec{v}$ is the semantic vector for a keyword in a music-physics lexicon, one could represent the domain of origin using a two-dimensional vector $\vec{\chi}$, with $\vec{\chi}=[1,0]$ for a music keyword, $\vec{\chi}=[0,1]$ for a physics keyword, and $\vec{\chi}=[1/\sqrt{2},1/\sqrt{2}]$ for an integrated keyword. Then, the original semantic vector could be augmented with a direct product $\vec{v} \otimes \vec{\chi}$, encoding information about disciplinary origin in a slightly expanded semantic space.

At the end of an unsupervised process, one could use lexical categories to aid in the interpretation of the emergent topics by categorizing their disciplinary weighting. For example, in the Physics of Music course studied here, a topic related to standing waves would likely be heavily weighted with physics keywords, while a topic related to spectral analysis of instruments would be closer to equally weighted with music and physics keywords. The network analysis techniques applied in this paper could then help explore how instances of these topics integrate concepts from these categories. This interpretive support aligns with computational grounded theory's emphasis on iteration between human interpretation and computer-based categorization \cite{tschisgale2023integrating,nelson2020computational}.

These benefits offered by lexical methods can be realized at scale, with the lexicon maintaining a consistent size and workload with increasing corpus size. We have demonstrated what insights lexical methods can find using a small dataset that we understand well from experience, but one could now take this lexicon and apply it to another corpus related to music-physics integration with an arbitrarily large number of artifacts of any textual length. This process can be repeated for other course contexts or theoretical models, using a small, well-understood corpus to construct a lexicon that can be applied to a larger corpus.


\section{Limitations\label{sec:limitations}}

In order to quantify the students' use of disciplinary keywords, we chose to use the Jupyter notebook's markdown cell as our unit of analysis. This choice ignores ``heavier'' usage of a keyword within a given markdown cell, which might indicate the relative importance a student places on a given keyword. However, we felt that counting at the sentence level or individual word level seemed too granular for our purposes, and could make the results dependent on a student's writing style.

We are also limited by the sample size of this course, with only 10 students. A larger sample size could permit the use of bootstrapping procedures \cite{rosvall2010mapping} to establish error estimates on our betweenness values. Similarly, the limited number of writing assignments collected does not permit a more systematic test of our interpretations of the results based on our assessment of an assignment's level of epistemic agency. However, our sample does suffice for a proof-of-concept of the type of insights that lexical methods can offer.

We found the process of keyword categorization to be straightforward and that we were able to resolve any ambiguities by referring to the context in which they were used. However, this process does rely on our interpretation of music terminology with which we are familiar but not experts. If categorizations were carried out differently, it would alter results like the category integration frequency in Figure \ref{fig:CL-content} and the music-physics betweenness values in Figure \ref{fig:music-physics-bewteenness}. Following Tschisgale et al.'s call for researchers to share their data and analysis scripts \cite{tschisgale2023integrating}, we make our lexicon, corpus, and Python scripts available at \cite{git}.

On a similar note, we make no claims about the degree to which students used these keywords appropriately. It seems reasonable to presume that these music majors mostly used their music keywords appropriately. However, students are generally known to misuse a physics keyword while referring to an underlying concept. For example, in the markdown cell from one student's CL1 that we quoted in Section \ref{sec:cl1_example}, they explained to their reader that ``the higher the frequency... the faster the speed of the wave.'' By ``faster speed,'' they might be mistakenly referring to the speed of the wave along the string (which is constant with respect to frequency), or they might be referring to how quickly an antinode must oscillate vertically to achieve a given frequency (thereby equating ``speed'' as a general concept with frequency).  The first author (who taught the course) reviewed the list of integrated and physics keywords and did not recall any conversations in class about misunderstandings of their meanings, and any significant misuses of terminology would have been addressed in the CL revision process. However, even if a student misused a physics or integrated keyword for the given physics context (such as might be the case in the quoted example), they were at least attempting to integrate music and physics concepts, pointing to a developing fluency in the new domain.

Finally, we should note that the music majors represented in this course are by no means monolithic: They were enrolled in various concentrations within the music major (instrumentalists and vocalists, music performance, music education, and music technology). Similarly, we did not collect data about their STEM backgrounds, which could exhibit significant variations across the population. Our study omits potential differences between these subgroups. For example, the subset of music technology majors might exhibit different facility with the programming tasks in the course discussed in \cite{lane2025computing}, or a student who plays two different types of instruments might have a more intuitive understanding of timbre than others. Such differences might influence how they integrate music and physics knowledge.

\section{Conclusions \label{sec:conclusions}}


We have demonstrated how lexical methodology can offer insight into structural aspects of students' written artifacts, particularly with how students integrate new physics knowledge with knowledge from a domain of interest. We have shown how this method provides a means of quantifying the epistemic agency in a writing assignment. Researchers can directly extend this methodology to other physics courses that integrate knowledge from physics and another domain, and they can choose lexical categories based on their theoretical model.

We have also outlined how lexical methods complement the textual analysis that STEM education researchers are increasingly exploring. Lexical methods offer a middle ground between supervised and unsupervised approaches, and can be run in parallel with textual analysis or can be directly integrated into the language models used in textual analysis. These insights are driven by the human researcher's categorization of the lexicon's keywords, the time for which does not scale significantly with the size of the corpus being studied. Future work with larger datasets will formalize approaches that combine lexical methods with textual analysis to more richly explore student thinking as documented in writing.

\section{Acknowledgments}

We are grateful to the Physics of Music students for their participation and engagement in the course. We thank the PER@UNF group for feedback during the development of our methodology, and Tor Odden for extensive commentary and encouragement in the development of this manuscript. This work is supported by the University of North Florida Summer Undergraduate Research Fellowship.

\bibliography{apssamp}

\end{document}